\documentclass[useAMS,usenatbib,usegraphicx]{mn2e}

\title[Comets populations around Sun-like stars]
{Debris discs and comet populations around Sun-like stars: the 
Solar System in context}
\author[J. S. Greaves et al.]{J. S.
Greaves$^{1}$\thanks{E-mail: jsg5 at st-andrews.ac.uk} \& M. C.
Wyatt$^{2}$\\
$^{1}$SUPA, Physics \& Astronomy, University of St Andrews, North
Haugh, St Andrews, Fife KY16 9SS, UK\\
$^{2}$ Institute of Astronomy, University of Cambridge,
Cambridge CB3 0HA, UK
}

\begin{document}

\date{Accepted 2009. Received 2009; in original form 2009}

\pagerange{\pageref{firstpage}--\pageref{lastpage}} \pubyear{2009}

\maketitle

\label{firstpage}

\begin{abstract}

Numerous nearby FGK dwarfs possess discs of debris generated 
by collisions among comets. Here we fit the levels of dusty 
excess observed by Spitzer at 70~$\umu$m and show that they 
form a rather smooth distribution. Taking into account the 
transition of the dust removal process from collisional to 
Poynting-Robertson drag, all the stars may be empirically 
fitted by a single population with many low-excess members. 
Within this ensemble, the Kuiper Belt is inferred to be such 
a low-dust example, among the last 10~\% of stars, with a 
small cometary population. Analogue systems hosting gas giant 
planets and a modest comet belt should occur for only a few 
per cent of Sun-like stars, and so terrestrial planets with a 
comparable cometary impact rate to the Earth may be uncommon. 
The nearest such analogue system presently known is HD~154345 
at 18~pc, but accounting for survey completeness, a closer 
example should lie at around 10~pc.
 
\end{abstract}


\section{Introduction}

Impacts pose a hazard to life on Earth, especially in the 
case of an event energetic enough to destroy much of the 
surface crust and oceans \citep{zahnle}. The most destructive 
impacts are associated with comets rather than asteroids, as 
the former hit at high speeds when infalling from the Kuiper 
Belt region \citep{sandra}, but the present-day cometary 
impact rate is low, largely because most of the primordial 
bodies have been dispersed \citep{morbidelli}. This is 
thought to have occurred in the first Gyr of the Sun's life, 
when Jupiter and Saturn crossed a mean motion resonance, 
destabilising the gas giants so that Saturn had close 
encounters with Uranus and Neptune. The expansion of their 
orbits perturbed the primordial Kuiper Belt \citep{gomes}, 
producing the Late Heavy Bombardment of the Earth at around 
700~Myr, after which the comet population was greatly 
depleted and the impact frequency has been much lower. 
However, the presence of the gas giants is still significant 
today, as dynamical interactions can bring comets into the 
inner Solar System -- \citet{horner} have recently shown that 
the impact rate on the Earth would vary significantly if the 
mass of Jupiter were different.

\begin{figure*}
\label{fig1}
\includegraphics[width=87mm,angle=270]{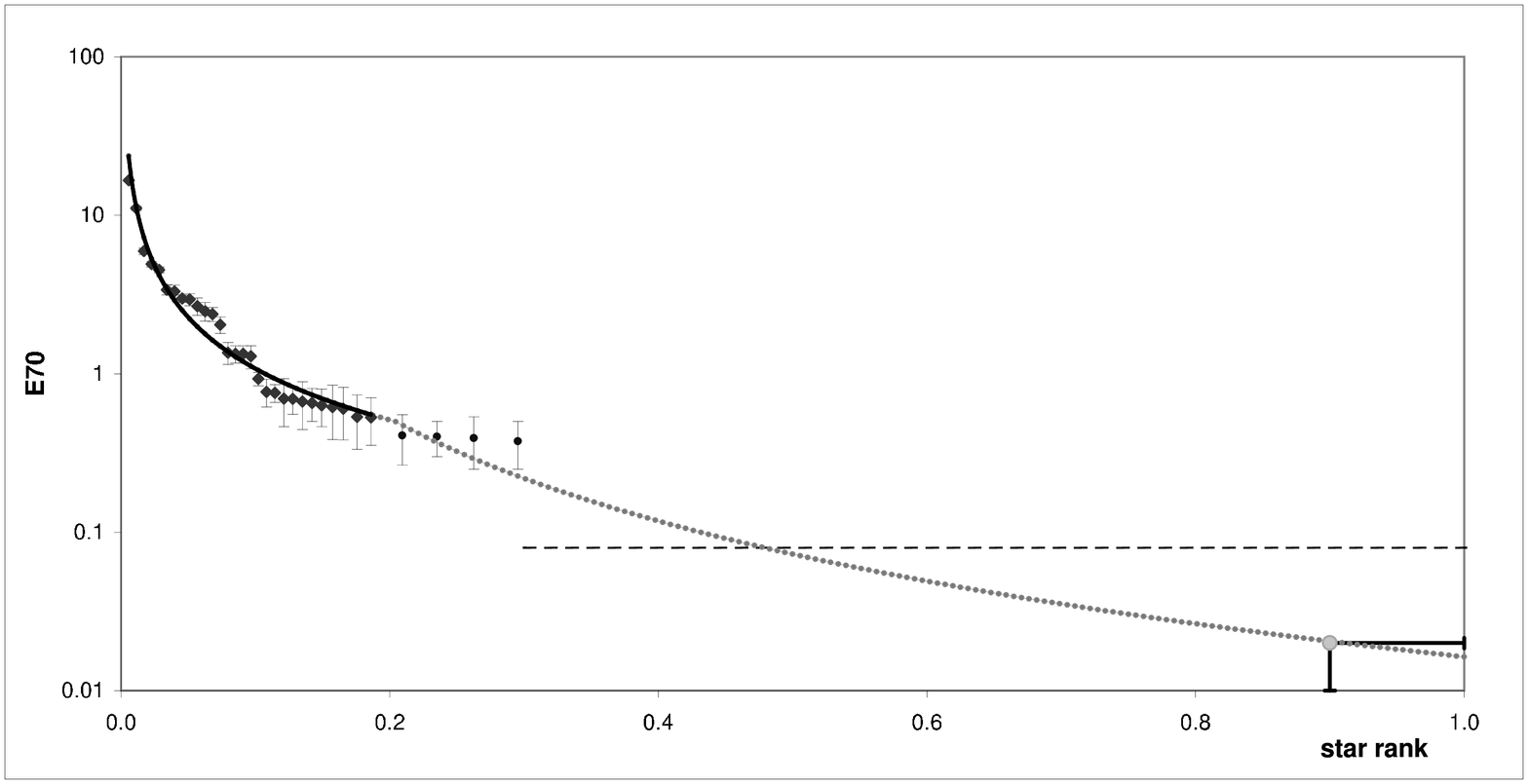}
\caption{Sun-like stars in order of decreasing 70~$\umu$m 
excess (y-axis). The rank on the x-axis is the star's number 
(e.g. 1 for the most dusty) divided by the number of stars 
surveyed. Upper limits are not plotted. The diamonds show 
contiguous debris detections while the four points plotted 
with circles have intervening systems with only upper limits in 
$E_{70}$; the spacing of these sparse detections is derived as 
described in section 3.1. The black curve is a power-law fit 
for stars of $E_{70}$ down to 0.5, excluding the four sparse 
detections of lower excess. The grey line is an extrapolation 
to lower excesses including dust removal theory (section 3.4), 
not a fit to the four circle symbols. The dashed line shows 
the $2.5 \sigma$ upper limit to the net $E_{70}$ for stars 
where detections beyond $x = 0.3$ could be made (section 3.2). 
The circle symbol (far right) is the position of the Sun in 
the ensemble, derived from its maximum estimated dust level 
(section 4.1) and plotted at its corresponding lowest rank on 
the grey curve. The bars indicate that the Sun may be less dusty 
and so of higher rank.
}
\end{figure*}

Thus, in extrasolar systems we may expect that the number of 
comets and the architecture of the giant-planet system will 
strongly affect the impact rate on any terrestrial planets 
present. Direct evidence of impacts in the inner systems 
comes from detections of dust grains at temperatures of a few 
hundred Kelvin, representing break-up debris from parent 
planetesimals. Such detections via a mid-infrared excess 
signal above the stellar photosphere are rare, but this does 
not imply that inner system planetesimals and planetary 
impacts are sparse, as the debris lifetime at a few AU is 
short \citep{wyatt05}. Less directly, we can examine the 
far-infrared debris signatures to assess the numbers of 
bodies colliding within the cool outer comet belts, and 
consider the influence of gas giants in perturbing comets 
inwards to where they threaten terrestrial planets.

Here we use the results of recent volume-limited surveys of 
Sun-like stars with Spitzer by \citet{fgk2} and 
\citet{simtpf} to study the dustiness of Solar-analogue 
systems, via the predominantly 70~$\umu$m detections of 
excess above the stellar photospheres.  Various analytical 
models, as reviewed by \citet{wyatt08}, predict far-infrared 
dust luminosities as functions of initial size and mass of 
the planetesimal disc and the age of the host star. Thus the 
excesses detected by Spitzer can be related to the ensemble 
of populations of comets per star. We can then place the 
Sun's Kuiper Belt, for which there are measurements of the 
numbers and distribution of comets and dust particles, in the 
context of similar stars. For example, other systems are 
known that are hundreds of times dustier than the Solar 
System; dustier at greater age than the Sun; or have much 
larger debris belts.  Viewed externally as a debris disc, the 
Kuiper Belt would be faint, a result which motivates this 
study to answer the question of whether the planet and 
planetesimal population around the Sun is unusual, and 
whether this has any role in impacts on the Earth and the 
evolution of life.

\section{Data}

Two major unbiased surveys of nearby Solar analogues have 
been made with Spitzer, with debris detection rates of 
$\approx$15~\% at 70~$\umu$m, for fractional excesses of a 
few tenths or more above the photosphere \citep{fgk2,simtpf}. 
The FGK targets in these two samples are complementary and 
extend out to $\sim 30$~pc from the Sun; a further survey 
\citep{kim} will complete much of this volume. We neglect 
here results from the FEPS project \citep{feps,feps2,meyer} 
where stars were selected by age, and so there is a trend of 
lower sensitivity for younger and thus rarer and more distant 
stars. The published nearby-volume samples also have some 
biases that are inherited from the goals of the original 
proposals, with more luminous stars, hosts of giant planets 
and single stars over-represented compared to their true 
proportions among Sun-like stars.

Only 70~$\umu$m data of high quality from \citet{fgk2} and 
\citet{simtpf} are discussed here, tracing cool dust at tens 
to hundreds of AU from the host star. The photospheric signal 
was not detectable with 3-sigma confidence in all the stars, 
so there is a potential bias where faint stars rise above the 
detection threshold of {\it total} flux only in cases of 
large excess. To avoid this, we selected only the objects 
with $F_*/\delta F > 3$, where $F_*$ represents the 
photosphere and $\delta F$ is the noise in the flux 
measurement. This selection includes 176 stars from the two 
combined surveys, after also eliminating a number of M-dwarfs 
from \citet{simtpf}. The excesses are then defined as $E_{70} 
= F_{dust}/F_* = (F_{total} - F_{*})/F_{*}$.  All further 
analysis uses the fluxes and errors listed by 
\citet{fgk2,simtpf} along with their other tabulated 
quantities, such as estimates for stellar ages.

The survey papers identify debris systems as those with 
$E_{70}$ at $3 \sigma$ confidence, i.e. $\chi_{70} = 
F_{dust}/\delta F \geq 3$, here comprising 27 stars. Here 
we add 6 more candidate systems of slightly lower 
significance. A tabulation of all 176 $E_{70}$ values, in 
order of $\chi_{70}$, shows that {\it negative} outliers 
extend down to $-2.5 \sigma$, and this distribution is 
quite symmetric, with 70 stars of $0 \leq \chi_{70} \leq 
+2.5$ versus 73 of $-2.5 \leq \chi_{70} < 0$. Further, this 
$\chi_{70}$ distribution is close to a Gaussian centred on 
a null value, with mean and sigma of -0.03 and 1.2 
respectively. We thus adopt here as candidate debris 
systems all those with $> +2.5\sigma$ significance, and 
estimate that the chance of a false positive is $\la 
1.5$~\% per star, given that $< 1$ more extreme outliers 
appear among 73 negative results.  With this new selection 
criterion, there are 33 systems identified here as debris 
hosts\footnote{Those of 2.5-2.9$\sigma$ confidence are HD 
39091, 55575, 69830, 114613, 173667 and 222237; HD~69830 is 
a confirmed debris system with shorter-wavelength excesses 
\citep{fgk2}.} , or 19~\% of the stars observed.

\section{Analysis}

\subsection{Detected population}

\begin{figure}
\label{fig2}
\includegraphics[width=50mm,angle=270]{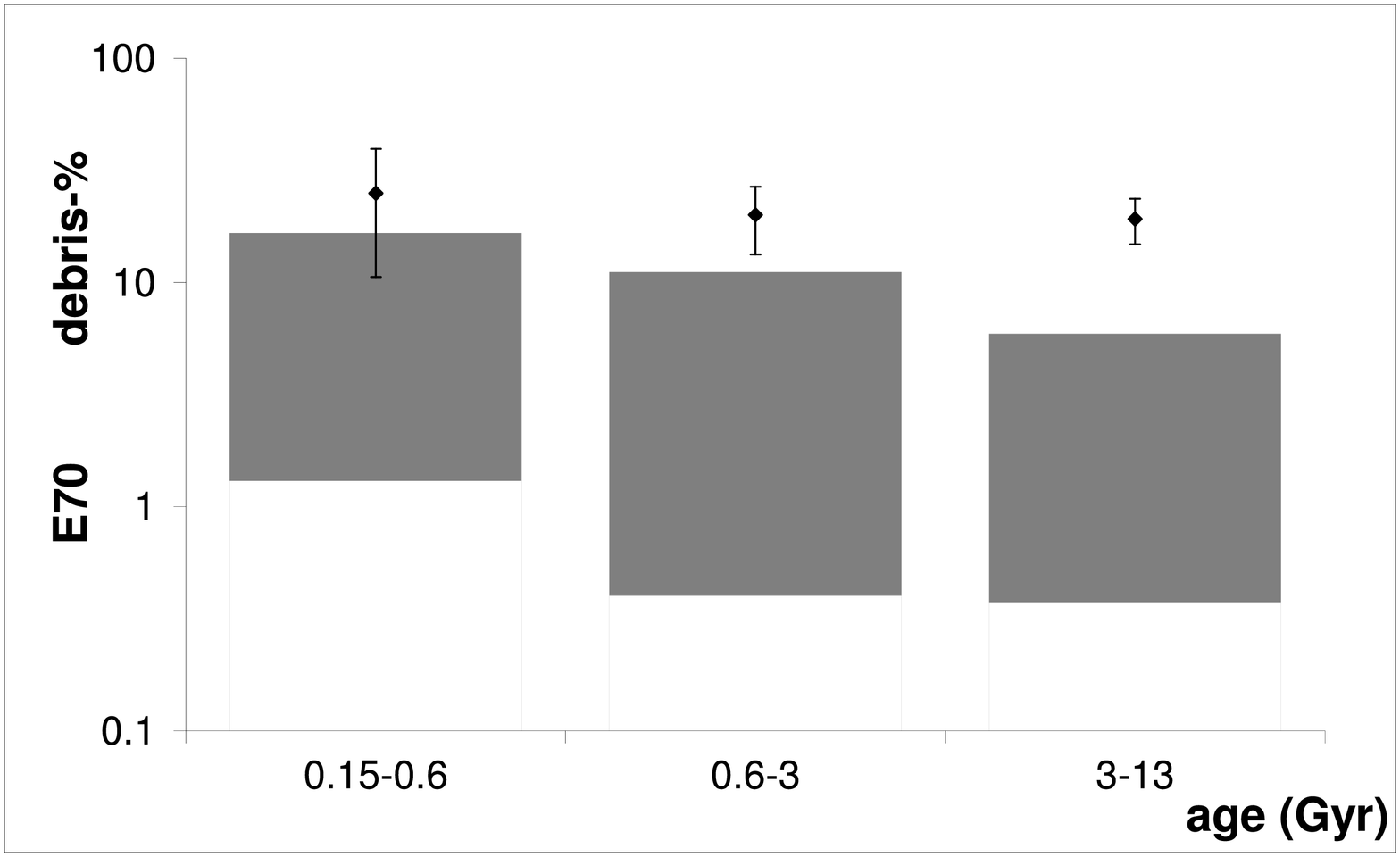}
\caption{Debris properties of the sample, in logarithmic 
age bins. The grey bars show the minimum-to-maximum $E_{70}$  
within each bin, and the points (and Poisson error bars) 
indicate the percentage of stars with debris. The number of 
systems detected (observed) in each age bin from left to right 
is 3 (12), 9 (45) and 19 (99), with 18 stars having no 
tabulated ages. 
}
\end{figure}

Figure~1 plots the 70~$\umu$m excesses of the stars where 
debris is identified, ordered by a fractional `rank'. The rank 
is formulated as the number of the star in order of decreasing 
$E_{70}$, divided by the total number of stars (176). It is 
equivalent to the fraction of stars with equal or higher 
$E_{70}$ than the current plotted point. However, beyond a 
rank of 0.1, not all the stars were observed to sufficient 
depth to detect the corresponding $E_{70}$ of approximately 
0.8 and below, if such excesses are present. This tendency is 
in fact implicit in the initial selection, as $\delta E_{70} = 
\delta F/F_*$ and so where $F_*/\delta F$ approaches 3, only 
$E_{70} > 0.83$ can be detected with $2.5 \sigma$ confidence. 
In this regime, there are four debris detections (circles in 
Figure~1) that are interleaved with upper limits of similar 
$E_{70}$, but the ranks of the non-detections can not be 
determined as the true debris levels are unknown. However, 
plotting these four points as contiguous ranks causes an 
artificial downturn in the trend of excess, because similar 
systems should occur in this region but were not actually 
observed deeply enough to be detected. Thus for these four 
points, estimated ranks are plotted, by increasing the step on 
the x-axis from 1/176, to 1/176 multiplied by a scale factor 
of (number of stars still to be counted) divided by (number of 
stars where the next $E_{70}$ could be detected). These stars 
with $E_{70} \approx 0.4$ were numbers 30 to 33 in order of 
decreasing dustiness, and their original ranks of 0.17 to 0.19 
(30/176 to 33/176) have been increased to 0.21 to 0.30.

The excesses observed range from 17 down to 0.4, although 
only about one in five stars was observed deeply enough to 
detect debris at this lower bound. The minimum detectable 
excess for Spitzer, where generally the dust and star lie 
within the same telescope beam, is set by uncertainty in the 
photospheric predictions plus errors in the 
shorter-wavelength data used for extrapolation. Dispersions 
in 24~$\umu$m data of around 5~\% are observed \citep{simtpf} 
and these are likely to be the minimum inherited errors when 
attempting to detect low levels of debris at 70~$\umu$m.

There is no strong trend of level or incidence of debris 
with particular stellar properties. For example, there is 
little trend with age, as already noted by e.g. 
\citet{fgk2,simtpf,gw}.  Figure~2 shows the sample divided 
into three broad bins each spanning a factor of $\approx 4$ 
in age (to minimise errors from poorly-dated objects). The 
detection rate is essentially flat with age over the entire 
main sequence, while maximum dustiness declines mildly with 
age -- a similar decline has been seen for 70~$\umu$m 
excesses of A-type stars \citep{su}. These trends agree 
with recent evolutionary models as summarised by 
\citet{wyatt08}, where each system's dustiness declines 
slowly with time as the parent population of colliding 
bodies is ground away. If the initial planetesimal discs 
have a wide range of radii and masses, then there is a wide 
range of excess values at any one age, and the detection 
rate is rather constant with time if the upper envelope of 
maximum dust flux is well above the detection limit 
\citep{gw}. 

Further, there are only weak trends associated with other 
stellar properties, such as luminosity, binarity and the 
presence of giant planets \citep{simtpf,fgk1,trillbin}. 
Figure~3 replots the debris hosts of Figure~1 with symbols 
denoting these properties, and no obvious trends are visible. 
In further analysis, all the stars are therefore treated as a 
single population.

\subsection{Trend in debris level}

\begin{figure}
\label{fig3}
\includegraphics[width=49mm,angle=270]{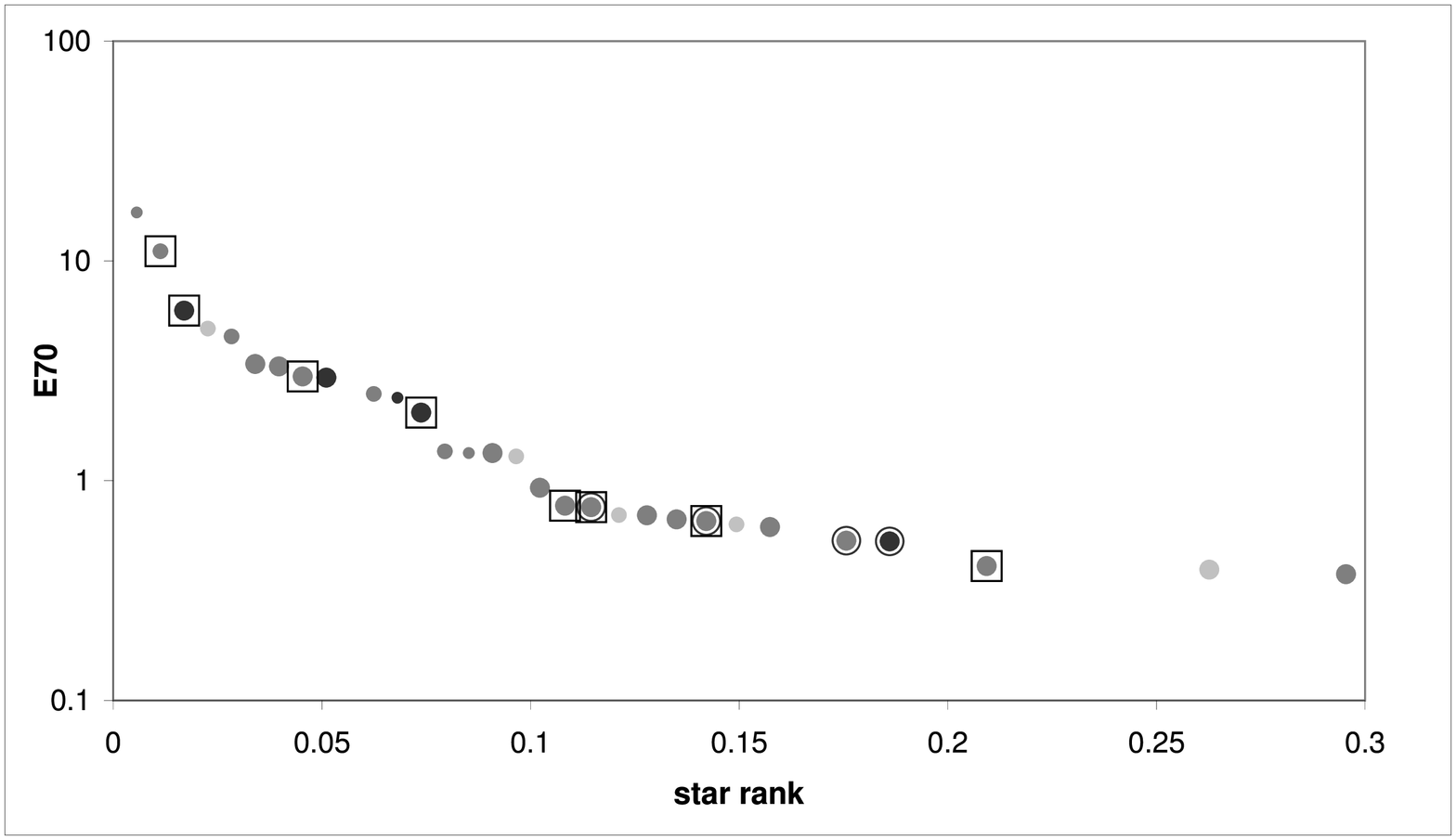}
\caption{Detected debris systems, showing stellar properties. 
Small, medium and large symbols correspond to the early, mid 
and late age-bins respectively of Figure~2, and light and dark 
grey shadings pick out the most and least luminous stars 
(early-F and K types respectively). Points within circles are 
systems hosting giant planets (from the Extrasolar Planets 
Encyclopedia) and those within squares have multiple 
stars (from the Catalog of Components of Double and Multiple 
Stars). 
}
\end{figure}

The ranked plot (Figure~1) shows that the excesses form a 
rather smoothly declining distribution, dominated by lower 
values of $E_{70}$. The steep slope can explain how the rate 
of far-infrared debris detection among Sun-like stars has 
risen with improvements in sensitivity with successive 
instruments, namely the IRAS, ISO and Spitzer missions. For 
example, a typical G-dwarf photosphere at 15~pc has a flux of 
approximately 30~mJy at 60~$\umu$m \citep{habing}, while the 
IRAS all-sky survey could detect signals of around 200~mJy 
\citep{wyatt05} and the corresponding limit for ISO was around 
90~mJy \citep{decin}. Thus IRAS and ISO could detect $E_{60} 
\sim 6$ and 3 respectively. Of the stars plotted in Figure~1, 
we would then expect about 3 bright examples to be discovered 
by IRAS, and HD~139664, 48682 and 109085 were actually found. 
Subsequently, ISO could have discovered $\sim 6$ less-dusty 
systems, while the sparse surveys actually undertaken added 
two discoveries\footnote{Five other fainter debris systems in 
Figure~1 were also in fact candidates from IRAS/ISO, where 
fainter excesses could be probed for various reasons, such as 
stellar proximity, or hotter photospheres or warmer dust than 
average.}, HD~30495 and 17925

\begin{figure}
\label{fig4}
\includegraphics[width=66mm,angle=270]{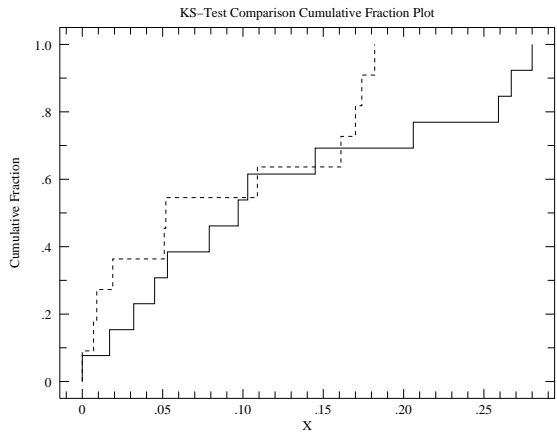}
\caption{Cumulative distributions of measured $E_{70}$ (X 
axis) for 24 upper-limit systems discussed in the text. These 
have small $E_{70}$ errors of $\leq 0.15$ and so the 
distribution is a guide to whether a net positive excess may be 
present. Positive values (solid line) are compared to negative 
values that have been multiplied by -1 (dashed line).  
}
\end{figure}

The large sensitivity improvement with Spitzer has more than 
tripled the discoveries of debris systems among these nearby 
stars, suggesting that more fainter examples remain to be 
discovered. However, debris levels in the presently undetected 
population are inferred to be small on average. \citet{ag90} 
suggested that a net excess of $\sim 0.05$ might be present, 
from IRAS 60~$\umu$m data for G-stars within 25~pc. The Spitzer 
70~$\umu$m data now explore this regime lying at the right of 
Figure~1, although individual excesses are still not detected. 
For example, there are 24 stars with Spitzer observations deep 
enough that excesses ranging from 3~\% to 37~\% could have been 
detected, if such dust levels were actually present in these 
systems. (Here we neglect the uncertainties due to calibration 
and just consider signals of 2.5 times the noise level.) These 
stars are part of the population at star-rank $>0.3$ in 
Figure~1, and none of them has a significant measurement of 
excess, but the data are deep enough that a tendency to mildly 
positive values can be searched for. In particular, stacking 
the data can show if there is evidence for an average non-zero 
dustiness that is not apparent in each individual noisy 
observation. Averaging the measured $E_{70}$ for these 24 stars 
we find $0.03 \pm 0.02$ for the mean value and standard error 
on the mean\footnote{The net value from all stars without 
debris detections is similar at $-0.01 \pm 0.02$ but is more 
influenced by unphysical negative excesses, down to $E_{70} = 
-0.7$. These may indicate uncertainties in the photospheric 
models, with the two lowest values corresponding to stars at 
the extremes of the spectral type range, for example.}. This is 
not a significant detection of net excess, so the suggestion of 
a typical excess of 0.05 by \citet{ag90} (at 60~$\umu$m) can 
not be confirmed. However, it does allow all systems to be 
slightly dusty, or some to have modest excess while others are 
dust-free, as well as the possibility that all are dust-less. 
Among these 24 targets there may in fact be a slight skew 
towards more positive $E_{70}$ measurements (Figure~4), but a 
Kolmogorov-Smirnov test finds a probability of only 54~\% that 
this positive tail are from a different (i.e. debris disc) 
population. Therefore there is as yet no conclusive result, but 
this does not rule out many low-dust systems. As a small excess 
is characteristic of the Solar System, attempting to infer the 
underlying population is useful to put the Sun in context.

\subsection{Single population hypothesis}

We first test the simplest possible hypothesis, that all
Solar-type stars have debris discs drawn from a single
distribution. A suitable function is sought that fits the
excesses, plus information in the low net excess for systems that
are not detected. The single-population hypothesis
is unlikely to explain all systems completely, but provides a
testable empirical approach. 

Figure~1 shows (black line) a best fit to the debris
detections with $E_{70} = 0.5$ to 17, given by 
\begin{equation}
E_{70} = 0.09 x^{-1.08}
\end{equation}
where $x$ is the star rank, and the correlation coefficient is 
0.98. The single-population hypothesis is thus a rather good 
empirical fit to the systems detected, in spite of the wide 
range of stellar ages, spectral types, etc. The regime at 
$E_{70} < 0.5$ was excluded as a different dust behaviour is 
predicted (see below), and includes only 4 detected systems. 

If the Equation~1 fit continued for all ranks up to 1, then 
the last star would have $E_{70}$ of 0.09, and the mean in the 
undetected population at $x > 0.3$ would be 0.16. If this 
extrapolation applied, more detections should have been made, 
given a 3-sigma threshold of $\sim 0.15$ set by uncertainties 
$\sim 0.05$ in the photospheric levels as discussed above. 
Also, the true mean is unlikely to be so high, as among the 24 
stars with the smallest excess-errors, there are only four 
objects with $E_{70}$ measurementts even as high as $\sim 0.25$ 
($2.5 \sigma$ upper limits $\sim 0.4$). This suggests a drop 
in dustiness among the majority population of stars.

\subsection{Dust generation}

The hypothesis above assumes one distribution for the amount of
dust per star, but given some range of initial circumstellar
disc properties, it is more likely that the quantity of comets
per star forms the underlying distribution. In this case, the
amount of dust generated needs to take into account the dust
removal processes, and as discussed by \citet{dominik}, this
depends on the number of colliding bodies. In massive discs,
the dominant destructive mechanism is collisional and $N_{dust}
\propto N_{comets}$, while in diffuse discs where the
Poynting-Robertson (light drag) effect removes dust, $N_{dust}
\propto N_{comets}^2$. This distinction between high- and 
low-mass discs, in the collisional and PR-drag dust removal 
regimes respectively, is also discussed by \citet{meyerppv}. 

\citet{wyatt05} has shown that the discs readily detectable in
the far-infrared are in the high-mass regime, but to extrapolate 
here to the whole population of stars, the dust generated in the 
low-mass cases must also be calculated. This transition occurs 
where the fractional luminosity of the debris is 
\begin{equation}
f_{crit} = L_{dust}/L_* 
= [2\times 10^4 (r/dr) (r/M_*)^{1/2}]^{-1}
\end{equation}
for disc radius and width $r, dr$ in AU and stellar mass $M_*$ 
in solar masses. It is obtained from the limit where PR-drag 
and collision lifetimes are equal, for grains where the 
radiative force is half the gravitational force 
\citep{wyatt05}. Using 11 resolved images of debris discs 
around Sun-like stars observed in the optical and 
submillimetre \citep{wyatt08,hd30495}, we calculated their 
values of $f_{crit}$. These lie in the range 1 to 10~$\times 
10^{-6}$ with an average of $5 \times 10^{-6}$. The 
corresponding 70~$\umu$m ratio $E_{crit}$ can then be 
calculated \citep{simtpf}, and for dust temperatures of 
40-80~K and stellar effective temperature similar to the Sun, 
$E_{crit}$ is on average 0.5 at 70~$\umu$m. Hence, if all 
Sun-like stars possess debris discs of similar scales to the 
resolved systems\footnote{Only 3 of the 11 resolved systems 
are in our present sample, and there could be a bias to larger 
systems where the aim is to make resolved images, although the 
resolved examples were not chosen for observation on such a 
priori expectations.}, then there should be a downturn in 
dustiness below $E_{70} \approx 0.5$, even if there is a 
single smooth distribution of number of comets per star.

The dotted line in Figure~1 shows this extrapolation beyond 
$E_{70} = 0.5$, assuming that the remaining systems are less 
dusty according to $E \propto N_{dust} \propto N_{comets}^2$ 
but continuing the same power-law distribution for 
$N_{comets}(x)$, i.e. of a form like Equation~1. In this 
case, the predicted mean $E_{70}$ is reduced from 0.16 to 
0.06, in better agreement with the estimate from the data of 
$0.03 \pm 0.02$ for all the stars at $x > 0.3$. Thus, the 
single-population hypothesis taking into account the different 
regimes of dust removal can match all the present 
observational results.

\subsection{Comet populations}

The number of comets per star can next be estimated from the 
dust excess, albeit via several scalings so the the final 
values are only approximate. To connect comets and excesses, 
we consider the $\tau$~Ceti debris disc, which is one of the 
nearby resolved examples and has $E_{60} = 0.75 \pm 0.15$ 
\citep{habing}. These ISO observations were made with a 
slightly wider and shorter wavelength-centred filter than the 
70~$\umu$m Spitzer/MIPS data, so we estimate a correction 
using stars detected by both ISO and Spitzer 
\citep{habing,fgk2}. The dust signals are similar at 60 and 
70~$\umu$m but the photosphere is fainter by a factor of $0.5 
\pm 0.1$ in the Spitzer data, from 6 G-dwarfs in common and 
with no debris. Hence, we adopt $E_{70} = 1.5$ for 
$\tau$~Ceti, while for the parameters of this system $E_{crit} 
\approx 0.85$, and so $E_{\tau {\rm Ceti}} \approx E_{crit} 
\times 1.8$.  Since $\tau$~Ceti has an estimated 
1.2~M$_{\oplus}$ of colliding bodies of radii $r \leq 25$~km 
\citep{tauceti}, under the assumption of a collisional cascade 
with incremental population $dN(r) \propto r^{-3.5}$, then 
$E_{crit}$ would correspond to 0.7~M$_{\oplus}$. However, 
models \citep[e.g.]{lohne} consider colliders up to $r = 
75$~km, so including this extra contribution raises the 
critical mass to 1.3~M$_{\oplus}$. Thus, with the assumption 
that the dust in all systems comes from a collisional cascade 
fed by 75km objects, the general scaling is
\begin{equation} 
E_{70}/0.5 = (M_{r \leq 75km}/1.3~M_{\oplus})^{\alpha}
\end{equation} 
where 0.5 is the generic value of $E_{crit}$, and $\alpha$ is 1 
for $E > E_{crit}$ and 2 for sub-critical discs. 

The mass in bodies of $r \leq 75$~km per star is then predicted
to range from 60 down to 0.2~Earth masses, with the detected
debris discs hosting $\ga 1$~M$_{\oplus}$ of comets. The shape
of the population is the same as the excesses fitted in
Figure~1, i.e. $M_{comets}$ approximately inversely 
proportional to $x$, so the most massive belts are rare. 
The probability that a star's comets exceed a given mass is 
found by inverting the rank-mass expression to give 
\begin{equation}
P(>M_{comets}) = 0.26 / M^{0.93}_{comets},
\end{equation}
applicable to the mass in bodies with $r \leq 75$~km.  The
highest mass appears realistic, for example it is comparable to
the 20-30~M$_{\oplus}$ of colliders inferred around the A-star
Fomalhaut \citep{wd}, which hosts a bright debris disc with
$E_{70} > 20$ \citep{karl}. The exact masses would be modified
if the comets do not follow a collisional cascade -- for 
example, \citet{hahn} fit the size-distribution for the Kuiper 
Belt as a double power-law with many small bodies but a 
depletion at sizes greater than 65~km. Adopting this 
distribution for the exo-comet belts would increase the 
fraction of the total mass that is in the collider population 
by a factor of about 2.5. 

%\begin{figure}
%\label{fig5}
%\includegraphics[width=58mm,angle=270]{versusmodel.ps}
%\includegraphics[width=60mm,angle=270]{muchdebrisFig4.ps}
%\includegraphics[width=83mm]{rank70_FGK.ps}
%\caption{Preliminary results for a synthetic disc population 
%compared to the Spitzer data. The hollow trangles are the 
%data from Figure~1 and the solid circles are model points 
%(see text). 
%}
%\end{figure}

\subsection{Models}

Although the fit used here is simply empirical, analytical models 
for evolving debris disc populations in fact produce similar trends. 
The models of \citet{wyatt07} have been applied to FGK stars 
surrounded by a range of planetesimal discs. These discs have masses 
drawn from a log-normal function and sizes from a power-law 
distribution, as in the previous model for debris discs of A-type 
host stars \citep{wyatt07}. The model systems are sampled as if 
observed at random ages and then ranked as in Figure~1. The general 
behaviour of a steep decline from a few very dusty systems down to a 
long tail of low-debris stars is easily reproduced, adopting 
planetesimal populations of a few Earth masses and discs extending 
out to a few tens of AU. Kains et al. (in prep.) will present the 
final model parameters fitting both 70 and 24~$\umu$m debris 
statistics among the Spitzer-surveyed FGK stars.

\section{The Solar System}

\subsection{Kuiper Belt properties}

We now compare the 70~$\umu$m excess of the Kuiper Belt (if it 
were viewed from outside) to the extrasolar values. The Solar 
System's far-infrared dust flux is unfortunately not well 
established, although \citet{landgraf} confirmed that Kuiper 
Belt dust does exist, via impacts onto the Pioneer spacecraft 
as they entered the outer Solar System. However the emission 
from this dust from an Earth viewpoint is a small perturbation 
to the much brighter Zodiacal flux. \citet{ag90} used IRAS 
scans to estimate that the peak Kuiper Belt flux is $3 \times 
10^{-8}$~W m$^{-2}$ sr$^{-1}$ at 60~$\umu$m (in a 40~$\umu$m 
bandpass), and the main belt on the sky is effectively 
$\approx 10^{\circ}$ wide\footnote{We include only the 
classical belt and neglect dust that could be generated over a 
wider belt by collisions among scattered disc objects. 
\cite{levison} suggest such an event created a recently 
discovered family of comets, but models are not well 
constrained by this one data point.} \citep{morbidelli}. 
Summing over this area and dividing by the bandpass to give a 
per-frequency-interval unit then gives a total flux of 
approximately 1~MJy ($\pm 0.5$~MJy from the scatter in 
different scans). Subsequent COBE data with a similar 
60~$\umu$m passband favour an upper limit at the lower end of 
this range \citep{backman}. We assume that the 70~$\umu$m dust 
flux will be similar, while the Solar photospheric emission 
would be 100 MJy if seen at the approximately 45~AU distance 
of the Kuiper Belt\footnote{This value based on the Sun's 
effective temperature was checked against a photospheric 
calculation for the MIPS/70 passband \citep{fgk2} for the 
`Solar twin' star HD~146233 \citep{soubiran}.}.

The Sun's excess at 70~$\umu$m is therefore $\approx 0.01 \pm 
0.005$ from IRAS, or possibly less from the COBE limit.  These 
values of excess for the Kuiper Belt are remarkably small. 
Allowing a $2 \sigma$ uncertainty so that $E_{70}$ can be as 
high as 0.02 would imply that $x \geq 0.9$. Hence, the Sun is 
one of the {\it least dusty} systems (Figure~1), or possibly 
not a member of the single population being tested in our 
hypothesis.

The total mass in Kuiper Belt bodies has been estimated at a 
few hundredths of an Earth-mass, from deep surveys finding 
objects as small as 25~km diameter \citep{bernstein}. The 
two-power-law model of \citet{hahn} can be used to estimate 
the total belt mass, adopting their generic albedo of 0.04. 
The Kuiper Belt would then sum to $\sim 0.08$~M$_{\oplus}$ in 
the $r \leq 75$~km regime (or somewhat smaller in a 
collisional cascade, or with more recently-adopted  
higher albedo). In contrast, in our single population 
hypothesis the lowest mass in colliders is $\approx 
0.2$~M$_{\oplus}$, and so the Kuiper Belt again appears to be 
an outlier, or at the extreme end of the range.

\subsection{Detectability of Kuiper Belt analogues}

An excess of order one per cent above the stellar photosphere
will be very difficult to detect around nearby stars, even with
the newly-launched Herschel mission. The PACS/70 chopping mode was
predicted to detect $\approx 3$~mJy sources with $5\sigma$
confidence in 1~hour, whereas the Sun if seen for example at
6~pc would have a photospheric signal of 135~mJy.  Such an
excess, at our Solar upper limit of 0.02, could thus in
principle be robustly detected in an hour. In practice, there
are only a dozen Sun-like stars within this distance, while
beyond it a disc of Solar System dimensions would span less
than three 5-arcsec telescope beams, and so uncertainties in
subtracting the blended stellar signal would be re-introduced.
However, Herschel will be able to explore somewhat larger
excesses very efficiently. For example, if 5~\% uncertainties
in stellar models allow the detection of $E_{70} \approx 0.15$,
Herschel is predicted to detect debris in 35~\% of 
systems (according to the grey line extrapolation in Figure~1), 
compared to about 20~\% of the stars observed with 
Spitzer. 

\section{Implications}

The single population hypothesis discussed above fits both the 
debris systems detected by Spitzer and the net upper limit for 
the remaining stars, once the switch of dust removal to 
Poynting-Robertson drag is taken into account. If the Sun is a 
member of this hypothetical population, it must be one of the 
least dusty systems, within the last ten per cent. The 
estimates of mass in colliding comets suggest the Kuiper Belt 
population may actually be a depopulated outlier, with only a 
few-hundredths of an Earth-mass in colliding bodies up to tens 
of kilometre sizes. Roughly a third of similar stars are 
predicted to host an order of magnitude more comets; 
therefore, implications for impacts on any terrestrial planets 
need to be considered.

Since the Sun is a normal mid-main sequence G-dwarf, these
results are surprising. Further, the primordial Kuiper Belt
should have been much dustier and a bright member of the
inferred single population. It is estimated that at least
10~M$_{\oplus}$ of rocky material was needed in order for the
Kuiper Belt bodies to form \citep{morbidelli}, and this total
belt mass would place the young Sun around $x \sim 0.1$ in
Figure~1. However, between the formation stage at $\sim
0.1$~Gyr and today, the Kuiper Belt should only have declined
about four-fold in dust luminosity \citep{lohne}, so its
present-day rank would be $x \sim 0.3$, inconsistent with the
observed rank of $x \ga 0.9$. 

The key to these results is the dispersal of many Kuiper Belt 
bodies, so that the population is now very depleted 
\citep[e.g.]{morbidelli}. The Nice model \citep{gomes} 
proposes that dispersal occurred when Jupiter and Saturn 
crossed their mutual 2:1 mean motion resonance as they 
migrated outwards; this event would have moved Uranus and 
Neptune onto enlarged eccentric orbits and consequently 
disrupted the comet belt. This model can explain ejection of 
many objects, and can be synchronized with the Late Heavy 
Bombardment of the Earth-Moon system at about 0.7~Gyr. 
However, the existence and timing of this event depend on the 
precise architecture of the system of gas giants 
\citep{thommes}. Such as dispersal of around 99~\% of the 
comet mass could readily shift the Solar System's excess to a 
low state and thus a high rank, as inferred here -- a process 
recently modelled in detail by \citet{booth}. However, they 
find that $< 12$~\% of Sun-like stars can have undergone such 
an event, as there is no global drop in dustiness at mature 
ages (Figure~2), so system clearing can not have been common. 
\citet{gaspar} infer roughly similar proportions, based on the 
small number of stars that could be undergoing clearout events 
in the 760-Myr old cluster Praesepe. Thus in the present 
context, although comet-clearing by giant planets  
could shift the ranks of individual stars to much higher 
values, the data imply that such stars are only a small 
sub-set. This may be neglected in the single population 
hypothesis, as the net excess at $x > 0.3$ is presently poorly 
defined, and so switching a few per cent of planet-hosting 
stars to a low- or zero-dust state would have little effect on 
the net debris estimate.

\subsection{Relation to giant planets}

From long-term trends in Doppler wobble surveys, 
\citet{cumming} estimate that $\approx 18$~\% of Sun-like 
stars host giant planets, of above about Saturn's mass and 
orbiting within 20~AU. Here we assume that any such planet 
orbiting within 3~AU would be disruptive, for example 
perturbing a terrestrial planet at $\sim 1$~AU into an 
unstable or eccentric orbit, or in some migration scenarios 
preventing a habitable planet from forming \citep{fogg}. 
Excluding this 8~\% of systems leaves 10~\% of stars hosting a gas 
giant at 3-20~AU, among which four out of every ten planets 
(generally above Jupiter's mass) are already discovered 
\citep{cumming}. A small correction should be made for 
approximately 1~\% of star systems found to host another 
lower-mass giant that is closer in, although this is less 
than the uncertainty of about 3~\% in the extrapolation to 
orbits as distant as 20~AU. The result is that $\approx 9 \pm 
3$~\% of Sun-like stars should host gas giants no closer in 
than 3~AU, and are therefore reasonably analogous to the 
Solar System, with one-third of this population already 
discovered. Hypothesizing optimistically that all these stars 
could also have an unknown ice giant planet like Uranus or 
Neptune, the minimum conditions are met for gas giant 
interactions that could end with disruption of a comet belt 
outside the ice giant (should such a belt exist).

It is presently uncertain how giant planet systems and comet 
belts are related. Few stars have both gas giants and debris 
discs \citep{g07,amaya}, and systems with both phenomena have 
similar excesses to non-planet-hosts \citep{bryden09}. However, 
there could be a class of systems with many infalling 
comets and frequent impacts on any terrestrial planet(s). One 
archetype is HD~10647 which has a very high $E_{70}$ of 50 and 
so an inferred $> 100$~M$_{\oplus}$ of colliders; this object 
is $> 1$~Gyr old \citep{liseau}, while the four dusty 
planet-hosts of Figure~3 are about mid-main sequence. 
Consequently, it could be very difficult for complex life to 
ever evolve in these systems with at least twenty times more 
colliders than the Sun, if gas giants perturb these bodies 
into the inner regions. (Energetic impacts more frequently 
than every $\sim 10$~Myr would probably hinder recovery of 
biodiversity \citep{krug}; however, creation of large heated 
impact basins could favour the appearance of simple 
thermophilic life \citep{abramov}.) On the other hand, whether 
there is any connection between {\it faint} debris levels and 
gas giants is very difficult to determine. There are 14 
planets hosts in our sample that do not have debris 
detections, and for these the mean $E_{70}$ is $0.05 \pm 
0.05$, not significantly different from $0.03 \pm 0.02$ 
estimated for null-debris stars in general.

\subsection{Analogue exo-Earths}

An analogue planet to our own is hypothesized to have a 
roughly similar impact rate. Impacts much faster than the rate 
experienced on Earth, of about one every 100~Myr for 10~km 
bodies, are likely to cause a serious extinction problem (e.g. 
halving the number of species after each event). However, a 
much slower impact rate could mean that life would stay at a 
very simple level, with no need to adapt to changes in 
environment. Here we consider a cometary population comparable 
to or lower than that of the Sun as similar, but also note 
that the slope of the inferred $E_{70}$ is shallow at large 
$x$, so a considerable population could be only a few times 
dustier (depending on the poorly known Solar $E_{70}$). With 
this basic hypothesis, we then need to link impact rates to 
the configuration of the giant planets and comet belts. 
Unfortunately, this is computationally expensive to study for 
many systems \citep[e.g.]{horner}. Modelling (Greaves, Jeffers, 
Horner et al., in prep.) will compare the Solar System, with 
rather few comets but several perturbing giant planets, to 
representative common extrasolar architectures, such as many 
more comets but fewer perturbing planets. In the interim, we 
can consider two cases at opposite extremes: one where debris 
and giant planets are unconnected phenomena, and one where gas 
giants beyond 3~AU are a hypothetical signpost to systems 
where comets are cleared out.

The results above then suggest that $\la 10$~\% of stars have 
as few comets as the Sun (based on its $E_{70}$ upper limit), 
while about 9~\% of stars host an innermost gas giant beyond 
3~AU. If these phenomena are uncorrelated, then the product 
of these rates implies that only up to $\sim 1$~\% of 
Sun-like star systems would be analogous to our own, in the 
sense of a small comet population concurrent with 
outer-system giant planets. In the alternative case where 
giant planets can help to enforce a low comet population, 
this fraction of stars rises to an upper limit of 9~\%, 
assuming further-out ice giants are also present and so 
interactions can clear out the comet belt. This estimate of 
1-9~\% for an analogue system may be optimistic, if the 
exo-Earth impact rate is critically sensitive to the 
configuration of gas giants; for example, the clearing of the 
Kuiper Belt required Jupiter and Saturn to cross a particular 
resonance, in the model of \citet{gomes}. Further, there may 
not be an exo-Earth around all Sun-like stars, or not within 
the Habitable Zone where water can be liquid on the surface; 
see \citet{raymond} for a model prediction that stars of $\ga 
0.8 M_{\odot}$ are the most suitable hosts. On the other 
hand, our estimate of only a few per cent of analogue systems 
is too pessimistic if in fact the single population 
hypothesis does not apply, and there are $\gg 10$~\% of 
Sun-like stars with little dust. This is not possible to test 
with the current data, but Herschel observations should soon 
be able to test whether many stars host little dust.

Adopting up to 9~\% as the preliminary estimate of the 
frequency of analogue systems allows us to estimate the 
distance to the nearest counterpart. In conjunction with the 
space density of Sun-like hosts, which for mid-F to late-K 
dwarfs is 0.01 pc$^{-3}$, the nearest analogue system should 
lie within 13~pc. The nearest {\it presently} known analogue 
candidate (in the sense of planet and comet content) is the G8 
star HD~154345 at 18~pc distance, which has no obvious Spitzer 
70~$\umu$m excess and hosts a Jupiter-like planet at 4~AU.  
However, if only about one-third of the systems with gas 
giants beyond 3~AU have so far been discovered, a completeness 
correction suggests that a nearer analogue could lie within 
about 12~pc, or slightly closer given that not all nearby 
Sun-analogues as yet have published debris data.

\section{Conclusions}

Systems roughly analogous to our own, hosting a modest comet
population and gas giants at a few AU or beyond, are inferred
to be rather uncommon. If Sun-like stars often host terrestrial
planets in the habitable zone, only a few per cent of these are
likely to have a similar debris environment. A few systems may
have catastrophic bombardment, even at mid-main sequence age,
if gas giants perturb some of their numerous comets into the
inner system. At the other extreme, the majority population
should be stars without giant planets but with many comets, for
which further modelling is needed to assess the rate of
planetary impacts. The closest presently-known system that is
roughly like our own lies at 18~pc, but another probably exists
with around 10~pc given the completeness of debris and giant
planet surveys so far. This result is encouraging for future
facilities aiming to study habitable exo-Earths, such as
DARWIN, TPF and ELT.

\section*{Acknowledgments} 

We thank the referee for comments that helped to clarify 
several lines of argument. JSG thanks STFC and SUPA for 
support of this work.

\bsp

\label{lastpage}

\end{document}